\documentclass[%
reprint,
superscriptaddress,
 amsmath,amssymb,
 aps,
]{revtex4-1}

\usepackage{graphicx}
\usepackage{dcolumn}
\usepackage{bm}
\usepackage{epstopdf}
\usepackage{hyperref}

\begin{document}


\title{Interaction between defect and skyrmion in nanodisk}

\author{Chengkun Song}%
\affiliation{
	Key Laboratory for Magnetism and Magnetic Materials of the Ministry of Education, Lanzhou University, Lanzhou, 730000, People’s Republic of China
}%

\author{Chendong Jin}%
\affiliation{
	Key Laboratory for Magnetism and Magnetic Materials of the Ministry of Education, Lanzhou University, Lanzhou, 730000, People’s Republic of China
}%
\author{Haiyan Xia}%
\affiliation{
	Key Laboratory for Magnetism and Magnetic Materials of the Ministry of Education, Lanzhou University, Lanzhou, 730000, People’s Republic of China
}%

\author{Yunxu Ma}%
\affiliation{
	Key Laboratory for Magnetism and Magnetic Materials of the Ministry of Education, Lanzhou University, Lanzhou, 730000, People’s Republic of China
}%
\author{Jianing Wang}%
\affiliation{
	Key Laboratory for Magnetism and Magnetic Materials of the Ministry of Education, Lanzhou University, Lanzhou, 730000, People’s Republic of China
}%

\author{Jianbo Wang}%
\affiliation{
	Key Laboratory for Magnetism and Magnetic Materials of the Ministry of Education, Lanzhou University, Lanzhou, 730000, People’s Republic of China
}%
\affiliation{Key Laboratory for Special Function Materials and Structural Design of the Ministry of Education, Lanzhou University, Lanzhou, 730000, People’s Republic of China}%

\author{Qingfang Liu}
\email{liuqf@lzu.edu.edu}
\affiliation{
	Key Laboratory for Magnetism and Magnetic Materials of the Ministry of Education, Lanzhou University, Lanzhou, 730000, People’s Republic of China
}%



\date{\today}

\begin{abstract}
Magnetic skyrmions are topologically protected stable magnetization configurations, which are expected to be a promising candidate as information carrier, while defect is inevitable and plays an important role on the stabilization and movement of skyrmion. In this paper, we investigated the influence of a point defect and a ring defect on the stabilization and dynamics of skyrmion in the nanodisk, where the defect are acquired by local modification magnetic material parameters. Considering the combined action of geometry confinement and pinning effect, we demonstrate the preferred skyrmion position as a function of defect type and distance between skyrmion and a point defect. We also show the confinement effect on skyrmion size in the presence of a ring defect due to circular symmetry. Finally, in the application of spin transfer nano-oscillator, we show that the skyrmion can be pinned or rotate in the nanodisk, the oscillator frequency can be modified in a large variation by the ring defect. These findings provide a complete understanding of interaction between skyrmion and defect in the confined geometry and may provide a good strategy for the design of skyrmion oscillators.
\end{abstract}

\pacs{Valid PACS appear here}
\maketitle


\section{Introduction}
Magnetic skyrmion is a topological whirl of magnetization configuration~\cite{nagaosaTopologicalPropertiesDynamics2013,buttnerDynamicsInertiaSkyrmionic2015}. Since it has been discovered in Bulk materials and ultrathin magnetic films~\cite{yu2010real,seki2012observation,yu2011near,jiang2015blowing,wiesendanger2016nanoscale}, the formation and stabilization of skyrmion are investigated theoretically and experimentally~\cite{buhrandt2013skyrmion,sapozhnikov2015two}. Two basic types of skyrmion are Bloch and Néel skyrmion stabilized by the competition of exchange interaction and Dzyaloshinskii-Moriya interaction (DMI)~\cite{bogdanov1994thermodynamically,everschor2018perspective}. Other skyrmion-like structures, such as antiskyrmion~\cite{nayak2017magnetic,koshibae2016theory,hoffmann2017antiskyrmions}, skyrmionium~\cite{zhang2016control}, biskyrmion~\cite{yu2014biskyrmion,peng2017real} and bimeron~\cite{shen2020current,gobel2019magnetic}, have been discovered in magnetic systems with anisotropy DMI or frustrated magnetic materials, etc. Due to its topological related properties where the skyrmion cannot been disturbed easily, small size down to manometer scale and rich driven method to displace, the skyrmion is promising in the application of spintronic devices as an information carrier~\cite{fert2013skyrmions,sampaio2013nucleation}. Many researches are focused on the generation and manipulation of skyrmions. Meanwhile, the defect in the real magnetic system is inevitable in these process. 

The skyrmions nucleation, stability, manipulation and velocity are strongly depends on material inhomogeneous and defects~\cite{hanneken2016pinning,romming2013writing,gonzalez2019analytical}. The material inhomogeneous include a vacancy or a hole in the magnetic film, and geometry confinement~\cite{lin2013particle,muller2015capturing}. Typical defects considered in magnetic films include the local variation magnetic parameters, such as exchange interaction~\cite{lin2013particle}, magnetocrystalline anisotropy~\cite{everschor2017skyrmion}, saturation magnetization~\cite{heinonen2016generation} and DMI~\cite{kim2017current}, which can be modified by employing the ion beam irradiation. The magnetic anisotropy can be controlled by applying a voltage~\cite{xia2018skyrmion,zhang2015magnetic}. Stosic et al. investigated the interaction between skyrmion and these kinds of defects in a Co monolayer on Pt~\cite{stosic2017pinning}. The preferred pinning location of skyrmion depends on the size and type of defect, where the pinning and scattering potential act on skyrmion appear by either increasing or reducing the localized magnetic parameters. In order to observe the skyrmion in magnetic film, defects are necessary in the pinning of the nucleated skyrmion and addressability.

Moreover, the mobility of skyrmion can be affected in the presence of a defect with particular geometry~\cite{kim2017current,upadhyaya2015electric}. By building a trap with line defects or adding defect barriers, the skyrmion can be guided along a desired trajectory without skyrmion Hall effect~\cite{muller2017magnetic,song2017skyrmion,castell2019accelerating}, which is paramount in the application of skyrmion based racetrack memory. The speed of the skyrmion along the line defect can also be increased. In addition to that, the skyrmion can be effectively slowed down or captured by a hole in the film~\cite{muller2015capturing}. This is beneficial for the precise positioning of skyrmion by artificially tuning the defect position. Pinned and unpinned states of skyrmion in the film can be used in the transistor~\cite{zhang2015magnetic}. There are some works which have considered the skyrmion dynamics in the random or patterned defect arrays~\cite{reichhardt2015quantized,reichhardt2015collective}, and the effect defects with different shape on skyrmion. However, the collective effects of material geometry and defects on the skyrmion are still lacking. Skyrmions behavior depends on the geometry~\cite{rohart2013skyrmion}, which can be utilized in film as racetrack memory or transistor~\cite{fert2013skyrmions,zhang2015magnetic}, or in nanodot as spin transfer nano-oscillators~\cite{zhang2015current}.

In this paper, we investigate the interaction between skyrmion and a point (ring) defect in the nanodisk by considering the combined action of geometry and defect. The interaction and preferred positions of magnetic skyrmion with different kinds of defects are presented, as well as the defect shape with point or ring, where the ring-shaped defect can be treated as an arrangement of point defects. The strength of attraction or exclusion for skyrmion depends strongly on the type and shaped of defects. With this in mind, we investigate the dynamics of skyrmion under electrical current in the presence of defect ring. Due to the confinement of the geometry of nanodisk, the skyrmion experiences a circle trajectory and oscillates in the nanodisk. The oscillation frequency can be tuned in a large range.

\section{Micromagnetic framework}
\begin{figure}
	\begin{center}
		\includegraphics[width=8cm]{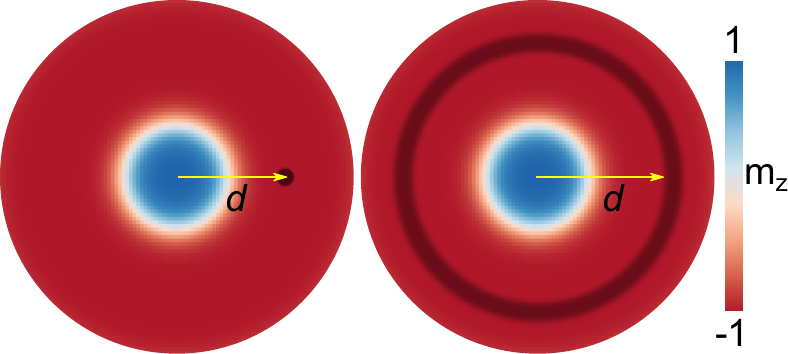} \caption{Schematic of the simulation system with a skyrmion in the presence of defects (black area), left panel is a point defect  with diameter 5 nm and right panel shows a ring defect with 5 nm width. $d$ is the distance between defect and nanodisk center.}
		\label{fig1}
	\end{center}{}
\end{figure}

We consider a ferromagnetic nanodisk of Co with perpendicular magnetic anisotropy on the top of a non-magnetic heavy metal layer of Pt, which provides substantial perpendicular magnetic anisotropy and interfacial DMI. In the competition of material parameters (exchange interaction, magnetic anisotropy and DMI), the skyrmion can be formed and stabilized in Pt/Co magnetic systems which have been proved in much literature~\cite{soumyanarayanan2017tunable,zhang2018direct,moreau2016additive}. The simulation system in this work is depicted in Fig.~\ref{fig1}, where an individual Néel skyrmion is stabilized in the center of the nanodisk as the initial state. Defects with point and ring shape due to the spatial variation of material parameters are represented by shadowed area in the nanodisk. In Cartesian coordinates, the plane of the nanodisk is the $xy$ plane, and the origin of coordinate is in the center of the nanodisk. The point defect is only positioned in +$x$ direction, and the distance of point and ring defect is represented by $d$. 

The magnetization of the system is described by classical spin $\mathbf{m}=\mathbf{M}/M_s$ with unit vector, and $M_s$ is the saturation magnetization. We describe the magnetic system using the free energy
\begin{equation}
E(\mathbf{m})=\int\left\{A[\nabla \mathbf{m}]^2 + K\cdot\mathbf{m}_z^2 + D\mathbf{m}\cdot(\nabla \times\mathbf{m})\right\}dr^2,
\end{equation}
which includes the exchange interaction $A$, magnetic anisotropy $K$ and DMI constant $D$. The dynamics of the system considering the current are simulated by solving Landau-Lifshitz-Gilbert equation numerically with micromagnetic simulation software Mumax3~\cite{vansteenkiste2014design}
\begin{equation} \label{eq:1}
\frac{d\mathbf{m}}{dt}=-\gamma\mathbf{m}\times\mathbf{H}_{\mathrm{eff}}+\alpha(\mathbf{m}\times\frac{d\mathbf{m}}{dt}) + \tau_{stt},
\end{equation}
where $\gamma$ and $\alpha$ are the gyromagnetic ratio and Gilbert damping, respectively. $\mathbf{H}_{\mathrm{eff}}$ is the effective magnetic field of the system, and $\tau_{stt}$ represents the spin transfer torque.

The thickness and radius of magnetic nanodisk considered in our simulation are 0.6 nm and 50 nm, respectively. In our calculations, we assume that an exchange stiffness constant of~\cite{sampaio2013nucleation,song2017skyrmion} $A=15\times10^{-12} \ \mathrm{J/m}$, a saturation magnetization of $M_{\mathrm{s}}=580\times10^3 \ \mathrm{A/m}$, a perpendicular magnetic anisotropy constant of $K_u=0.8\times10^6 \ \mathrm{J/m^3}$ and the DMI constant is $D=3.5 \ \mathrm{mJ/m^2}$. These parameters are varies in the defect area. The damping constant $\alpha$ is set as 0.3 in the static process and 0.02 for dynamics, and the temperature is considered as zero. 

\section{Effect of a point defect}
\begin{figure*}[ht!]
	\begin{center}
		\includegraphics[width=16cm]{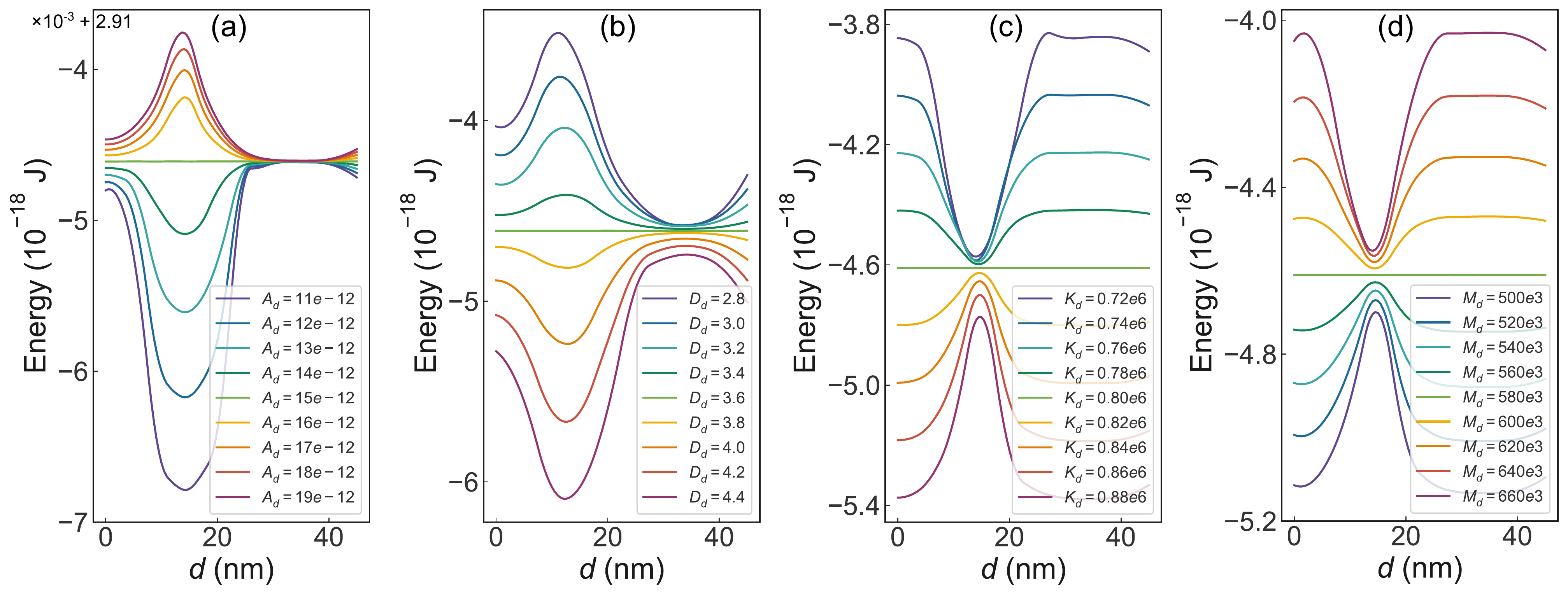} \caption{Total energy of the system as a function of the position of point defect with four kinds local variation of material parameters: (a) Magnetic exchange interaction ($\mathrm{J/m}$), (b) DM interaction ($\mathrm{mJ/m^2}$), (c) Anisotropy ($\mathrm{J/m^3}$) and (d) Magnetization ($\mathrm{A/m}$).}
		\label{fig2}
	\end{center}{}
\end{figure*}
We first consider the skyrmion behavior in the presence of a point defect with diameter 5 nm, which is regarded as a modification of material parameters in the point region with respect to the background of the system. Magnetic exchange, DMI, anisotropy and magnetization at the point defect area are represented by $A_d$, $D_d$, $K_d$ and $M_d$, respectively. It should be mentioned that a local variation of one parameter will not induce the changes in other interactions. Figure.~\ref{fig2} plots the energy of the system as a function of $d$ in the presence of four different kinds of point defects. The results show that the energy depends on the skyrmion-defect separation and local variation of a given magnetic parameter. By modifying the defect parameters and $d$, it is found that, in the presence of magnetic exchange and DMI defect, the energy experiences a large change with respect to the system energy without defects when $d$ is close to 15 nm. While the energy variations with $d$ reach the minimum value in the presence of magnetic anisotropy and magnetization defects when $d$ is close to 15 nm. It shows that the system energy increases when $A_d > A$ and $M_d > M_s$, and decreases conversely. While in the presence DMI or anisotropy defect, the energy decreases when $D_d > D$ and $K_d > K$. Then, we will figure out why a large energy variation exists when the distance between skyrmion and defect is about 15 nm. 

\begin{figure}
	\begin{center}
		\includegraphics[width=8cm]{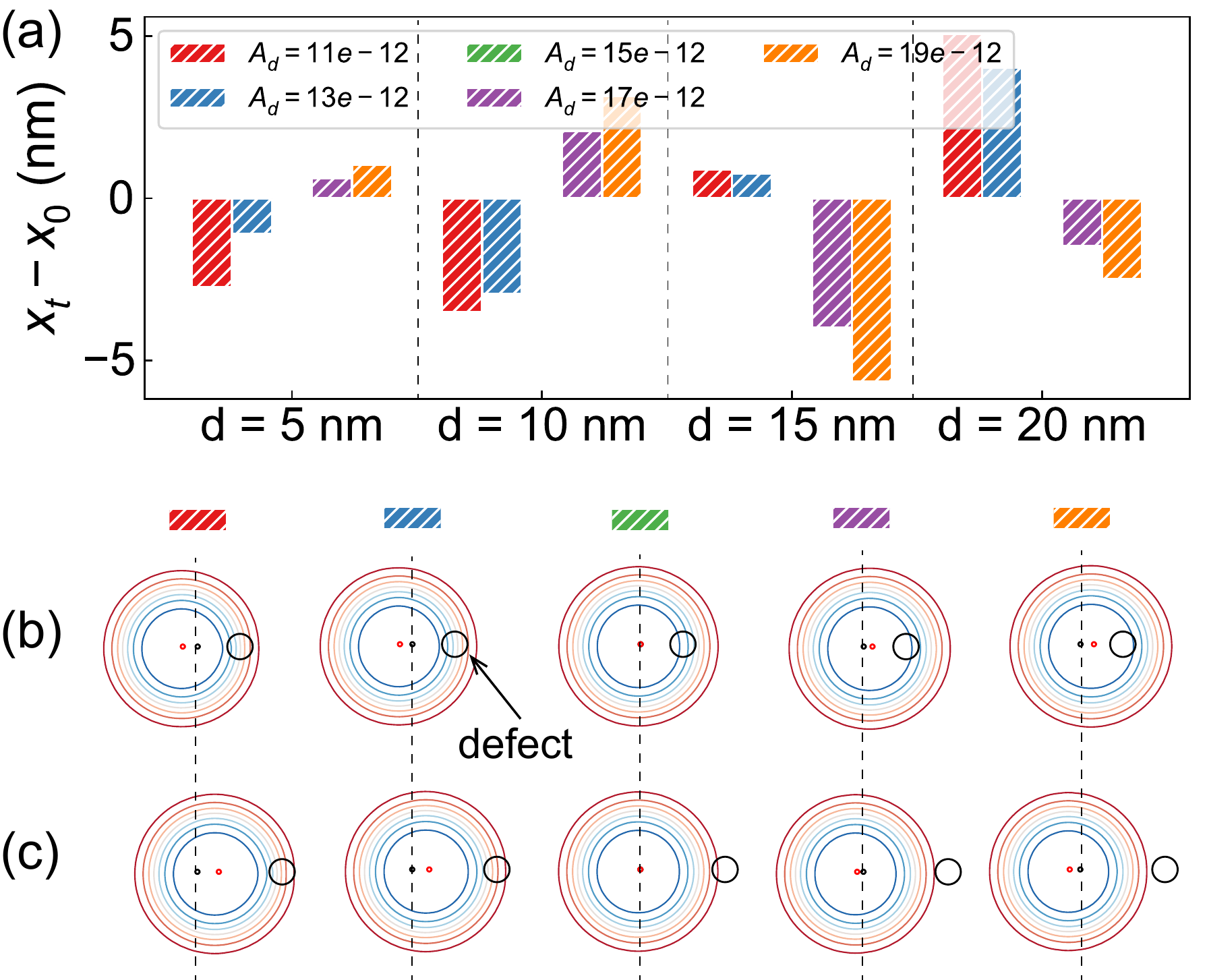} \caption{(a) Skyrmion displacement relative to the initial position with exchange point defect in different $d$. The corresponding position between skyrmion and point defect with (b) $d$ = 10 nm. (c) $d$ = 20 nm. Black and red points represent skyrmion initial and relaxed position, respectively.}
		\label{fig3}
	\end{center}{}
\end{figure}

An attractive or a repulsive potential acted on skyrmion with applying point defect, as have been pointed in ref.~\cite{stosic2017pinning,toscano2019building,fernandes2018universality}, that the skyrmion core moves towards or away from the magnetic defect due attractive or repulsive skyrmion-defect interaction. We show that the preferred skyrmion displacement due to local increase or decrease of $A_d$ is opposite. In addition, it depends on the skyrmion-defect distance $d$ with same $A_d$, as shown in Fig.~\ref{fig3}. The skyrmion displacement is characterized by distance between the topological center of relaxed state $x_t$ under defect and initial state $x_0=0$ nm, which is $x_t-x_0$. Figure.~\ref{fig3} (a) shows the skyrmion displacement in the presence of exchange defect at $d=5, 10, 15, 20$ nm, respectively. With decreasing $A_d$, the skyrmion is excluded by defect along -$x$ direction when $d=5$ nm and $d=10$ nm, and it is attracted along $x$ direction with increasing $A_d$. The relative positions of point defect (Black circle), initial skyrmion state (Black point) and relaxed skyrmion state (Red point) are shown in Fig.~\ref{fig3} (b) at $d=10$ nm. Whereas the skyrmion is attracted along $x$ direction with decreasing $A_d$ and excluded along -$x$ direction with increasing $A_d$ at $d=15$ nm and $d=20$ nm (Fig.~\ref{fig3} (c)). The exchange interaction favors parallel alignment of the spins, thus the skyrmion is pinned at the domain wall when a lower exchange interaction is considered ($A_d<A$). Conversely, the pinning is favored in single domain area ($m_z=1$ or $m_z=-1$) when a larger exchange interaction is considered ($A_d>A$). Depending on the skyrmion-defect distance $d$, the skyrmion is excluded in a smaller $d$ and attracted in a larger $d$ when decreasing exchange interaction $A_d<A$, where the defect area is overlapped with skyrmion domain wall part. In the case of $A_d>A$, the pinning position is favored in the skyrmion center part ($m_z=1$) and the skyrmion is attracted in a smaller $d$, and the relaxed position is favored outside the skyrmion ($m_z=-1$), thus the skyrmion is excluded in a larger $d$. 

\begin{figure}
	\begin{center}
		\includegraphics[width=8cm]{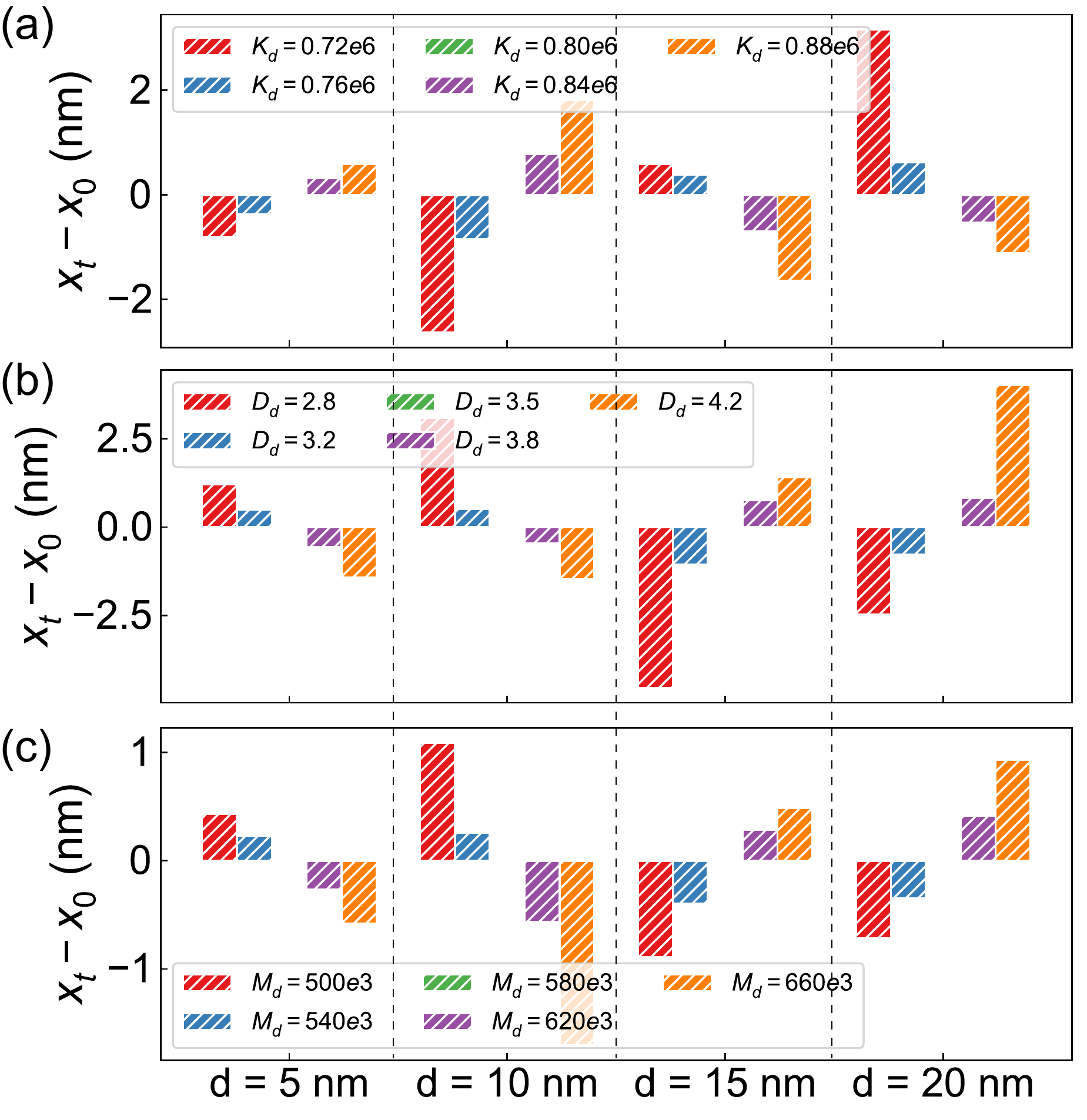} \caption{Skyrmion displacement relative to the initial position under a defect with (a) anisotropy (b) DMI (c) magnetization variation in point area at $d=5, 10, 15, 20$ nm.}
		\label{fig4}
	\end{center}{}
\end{figure}

A similar skyrmion favorable positions are found in the presence of an anisotropy point defect, as shown in Fig.~\ref{fig4} (a). With considering a point defect with lower anisotropy ($K_d<K$), the skyrmion is pinned at the domain wall, the skyrmion-defect interaction is repulsive in a smaller $d$ and attractive in a larger $d$, where $x_t-x_0$ is negative and positive respectively. While considering a larger anisotropy ($K_d>K$), the skyrmion is pinned at the single domain region, the skyrmion-defect interaction is attractive in a smaller $d$ and repulsive in a larger $d$, where the skyrmion is pinned in the center part and background of skyrmion, respectively. In the presence of a defect by varying DMI, which favors canting of the spins, the skyrmion is preferred to be pinned at the skyrmion domain wall part with a larger DMI ($D_d>D$), and pinned in the single domain area ($m_z=1$ or $m_z=-1$) with a smaller DMI ($D_d<D$). The skyrmion is excluded with negative displacement under a lower $d$ and larger $D_d$, and it is attractive and positive at a larger $d$ and lower $D_d$, where the displacement is opposite to the situation with exchange and anisotropy defects, as depicted in Fig.~\ref{fig4} (b). The skyrmion-defect interaction with considering magnetization defect ($M_d$) is analogous to the DMI defect, as shown in Fig.~\ref{fig4} (c). We have shown that an asymmetric pinning phenomenon appears for skyrmion in the presence of point defect, which can be attracted and excluded. The skyrmion displacement reaches maximum at $d$ close to 15 nm, which is in consist with the energy change shown in Fig.~\ref{fig2}. Above we have discussed an anisotropy point defect effect. Next, we will discuss the situation in the presence of an isotropic ring defect with circular symmetry.

\section{Effect of a ring defect}
\begin{figure*}[ht!]
	\begin{center}
		\includegraphics[width=16cm]{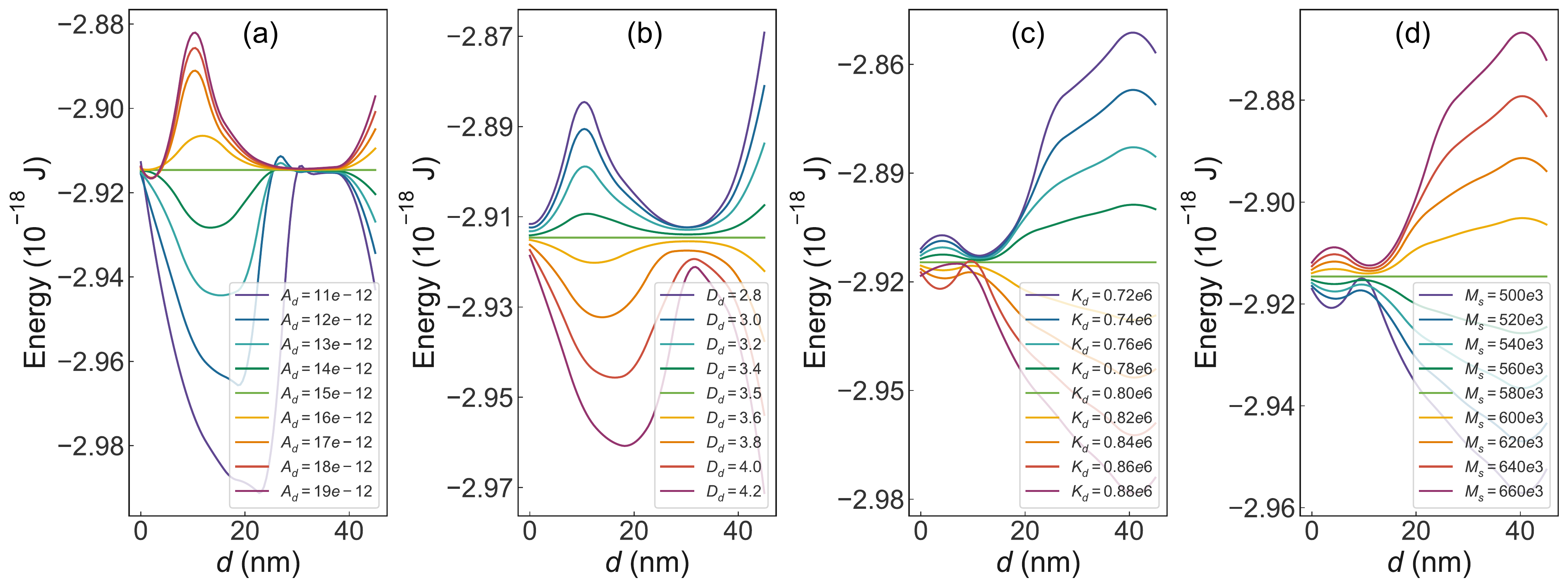} \caption{Total energy of the system as a function of the inner radius of ring defect ($d$) for four kinds of material parameters: (a) Magnetic exchange interaction ($\mathrm{J/m}$), (b) DM interaction ($\mathrm{mJ/m^2}$), (c) Anisotropy ($\mathrm{J/m^3}$) and (d) Magnetization ($\mathrm{A/m}$).}
		\label{fig5}
	\end{center}{}
\end{figure*}
Figure.~\ref{fig5} shows the system energy as a function of $d$ with decreasing or increasing local magnetic parameters in ring defect area. It shows that the energy increases with increasing $A_d$ and $M_d$, and decreasing $D_d$ and $K_d$, which is similar to that in the presence of a point defect. In the presence of exchange and DMI defect, the energy exhibits a large variation when $d$ is in the range of 15 nm and 20 nm, which is close to the skyrmion radius. With increasing $d$ to 35 nm, the energy variation is getting smaller, and it increases with $d>35$ nm due to the modification of nanodisk boundary. In the presence of anisotropy and magnetization defect, the energy variation of the system is small when $d$ is less than 15 nm, while it experiences a large change when $d>15$ and the variation is getting larger with increasing $d$. Depending on defect area, the energy variation under a ring defect is much larger than that in the presence of a point defect.    

\begin{figure}
	\begin{center}
		\includegraphics[width=8cm]{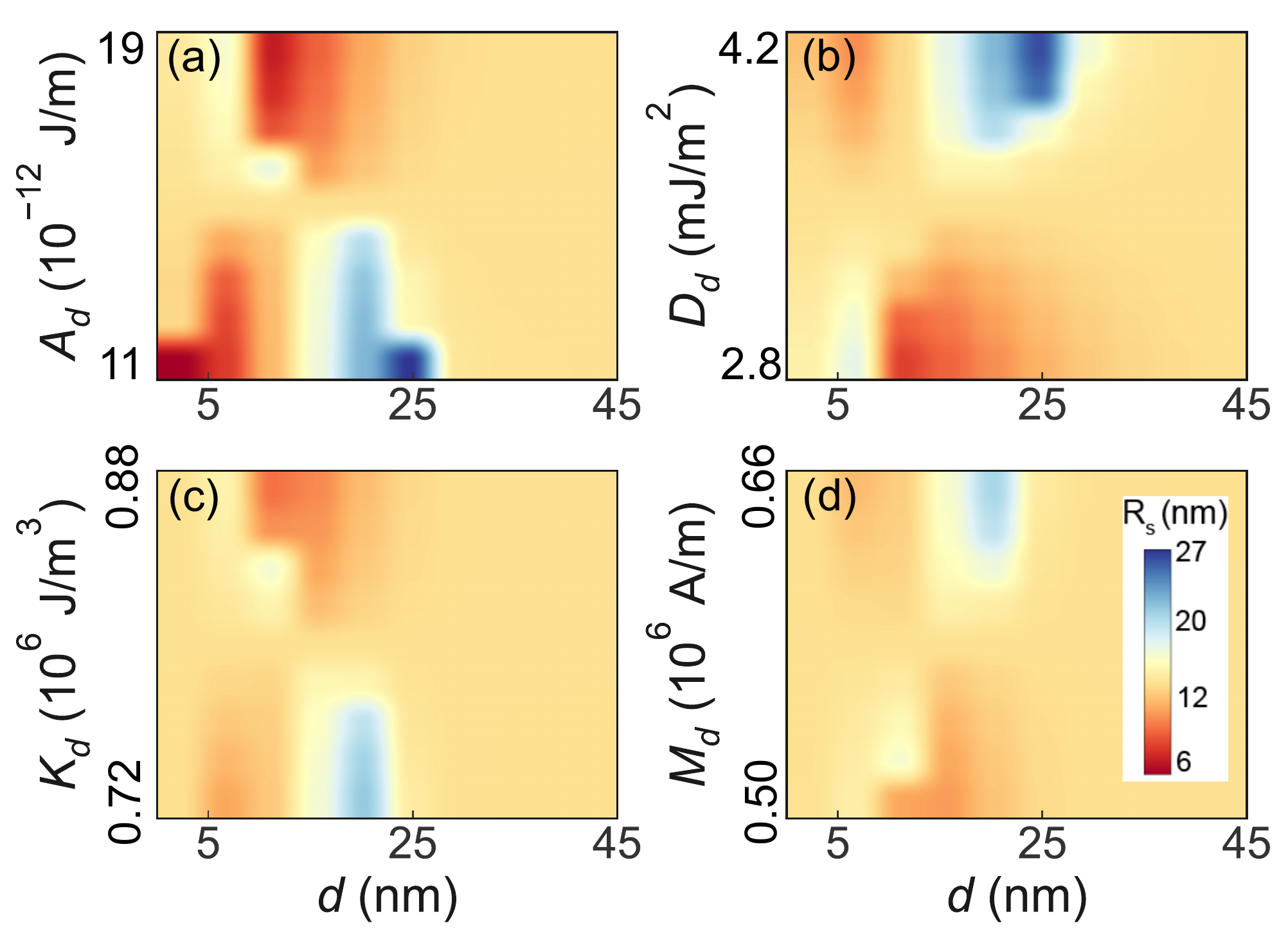} \caption{Skyrmion radius $R_s$ under a circle-like defect with different inner radius $d$ for four kinds of defect parameters (a) Magnetic exchange, (b) DM interaction, (c) Anisotropy and (d) Magnetization.}
		\label{fig6}
	\end{center}{}
\end{figure}

Due to the circular symmetry of a ring defect in the nanodisk, the skyrmion displacement is zero compared to that with a point defect. While the skyrmion size change is not negligible, which can be used to identify the skyrmion-defect interaction is attractive or repulsive. Figure.~\ref{fig6} shows the ring defect induced size change of skyrmion, we find that the skyrmion size experiences a large variation when $d$ is in the range of 0 to 30 nm. The skyrmion radius without defect is 14.8 nm (Yellow region). By decreasing $A_d$, the domain wall part are confined in the ring defect region, where the skyrmion size decreases with lower $d$ and increases at a larger $d$ until 25 nm, as shown in Fig.~\ref{fig6} (a). Whereas increasing $A_d$ induces the  decreasing of skyrmion size, for which skyrmion is extruded at the range of $10<d<20$ nm. This phenomenon can be understood by that the exchange favors parallel alignment of the spins. The anisotropy defect induced size change is analogous that in the presence of an exchange defect (Fig.~\ref{fig6} (c)). It is well known that the anisotropy favors spins normal to the nanodisk plane, thus a smaller $K_d$ induces an expansion of domain wall to ring area, and the skyrmion size changes with $d$. In a larger $K_d$, which can be treated as a potential barrier, and the skyrmion is extruded with small size. While in the presence of DMI defect, which favors canting of the spins, the size variation is opposite to the situation with $A_d$ ($K_d$). The ring with magnetization variation ($M_d$) induced skyrmion size change is similar to the case with DMI defect ($D_d$).

\begin{figure*}[ht!]
	\begin{center}
		\includegraphics[width=16cm]{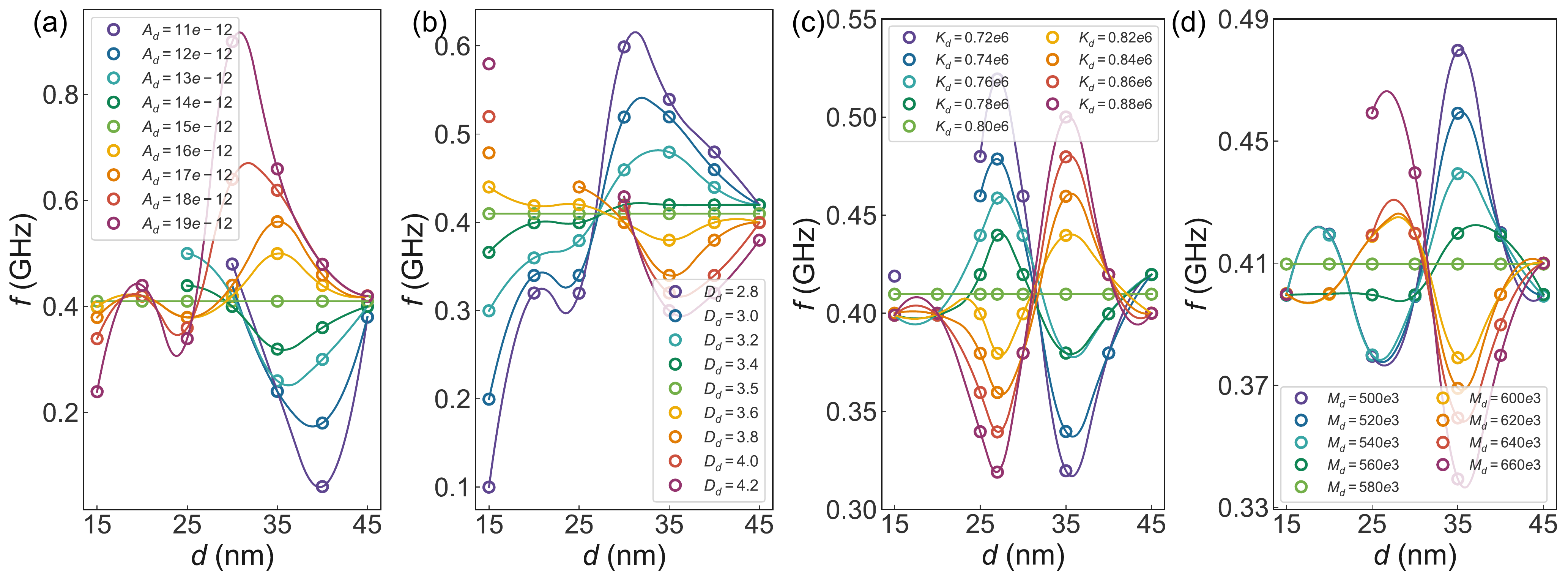} \caption{Skyrmion oscillation frequency $f$ as a function of the inner radius of ring defect with variation of four kinds of material parameters (a) Magnetic exchange interaction ($\mathrm{J/m}$), (b) DM interaction ($\mathrm{mJ/m^2}$), (c) Anisotropy ($\mathrm{J/m^3}$) and (d) Magnetization ($\mathrm{A/m}$).}
		\label{fig7}
	\end{center}{}
\end{figure*}
\begin{figure}
	\begin{center}
		\includegraphics[width=8cm]{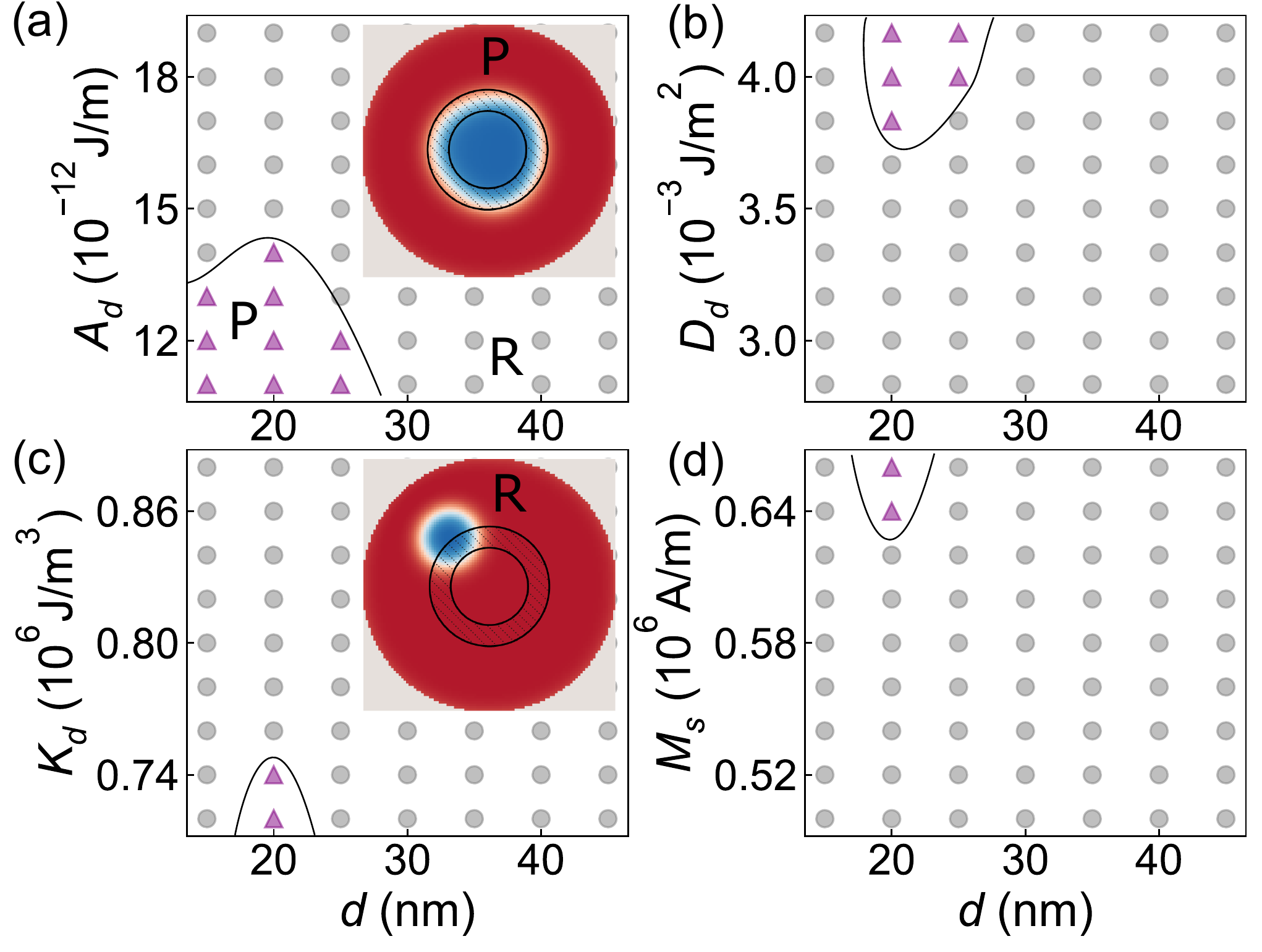} \caption{Phase diagram of skyrmion moving states for defect ring with varying (a) $A_d$, (b) $D_d$, (c) $K_d$ and (d) $M_d$, pinning state (P) and rotation state (R) are represented by purple triangles and gray circles, respectively.}
		\label{fig8}
	\end{center}{}
\end{figure}

 
Since we have demonstrated the static properties in the presence of a point defect and a ring defect by varying magnetic parameters, we next discuss the dynamics of skyrmion under perpendicular spin polarized current. We consider an electric current flows locally in the nanodisk center, where the current area is $2\pi r^2$ and $r=10$ nm and the current density is $j=5\times10^{11} \ \mathrm{A/m^2}$. Due to the spin transfer torque and boundary effect acted on skyrmion, the skyrmion exhibit a circular trajectory in the nanodisk, which can be used in spin transfer nano-oscillators (STNOs)~\cite{zhang2015current,garcia2016skyrmion}. Here, we investigate the oscillation frequency and pinning effect of skyrmion in the presence of a ring defect where material parameters are changed. Figure.~\ref{fig7} shows the skyrmion rotation frequency $f$ as a function of $d$ and material parameters in the ring defect area. We find that the frequency can be tuned in a large range. The rotation frequency is 0.41 GHz in the absence of ring defect. Considering a magnetic exchange ring defect ($A_d$), $f$ decreases with increasing $A_d$ at $d=15$ nm, while the skyrmion is pinned by the ring with the domain wall expanding in the ring region when increasing $A_d$, as shown in Fig.~\ref{fig7} (a). The spin transfer torque acted on skyrmion is not enough to conquer the pinning force, this phenomenon also occurs at $d=20$ and $25$ with increasing $A_d$. An obvious frequency change appears in the case of a large $d$, such as 30, 35 and 40 nm, where the frequency increases with increasing $A_d$ and decreases on the contrary. In the presence of DMI ring defect, a tremendous frequency variation can be modified at $d=15$ nm, for which increases as a function of $D_d$, as depicted in Fig.~\ref{fig7} (b). As increasing $d$, it shows a small modification of frequency with increasing $d$ close to 25 nm, and $f$ decreases at a larger $D_d$ until 35 nm, while it increases for smaller $D_d$ until 30 nm. This is due to the attraction and exclusion force between the ring and skyrmion for increasing and decreasing $D_d$, respectively. While for the defect ring with large $d$ away from the disk center, the interaction between them becomes weak and the skyrmion oscillation frequency shows small difference with the defect parameter variation. It is worth noted that the oscillation frequencies at some situations are not depicted at Fig.~\ref{fig7}, for that the skyrmion is pinned at the defect ring which represented by purple triangles in Fig.~\ref{fig8}. Figure.~\ref{fig8} shows the phase diagram of skyrmion moving states in the disk at different $d$ and parameter variation of defect ring, which are pinning (P) and rotating (R). In the case of skyrmion pinning, the domain wall part of skyrmion expands to the region of defect ring, thus the current can not drive skyrmion out to move. For smaller $d$, the skyrmion is easily pinned at the defect ring with decreasing $A_d$ and $K_d$, while the pinning effect arises when increasing $D_d$ and $M_d$. By comparing with the results shown in Fig.~\ref{fig6}, it is obviously that the defect ring plays an import role on the skyrmion size or skyrmion moving state when the radius of defect ring is comparable with skyrmion size. In the case of anisotropy defect, as shown in Fig.~\ref{fig7} (c), the oscillation frequency keeps almost unchanged with $K_d$ when $d$ is lower than 20 nm. By increasing $d$ to 27 nm, the difference between frequencies reach to a maximum, $f$ decreases with decreasing $K_d$. Conversely, it increases with increasing $K_d$ at $d=35$ nm. The same frequency modification but with opposite trend are depicted for magnetization defect, as shown in Fig.~\ref{fig7} (d), where the frequency increases and decreases with increasing the defect parameter $M_d$ when $d=27$ nm and $d=35$ nm, respectively. 

In order to figure out the influence of ring defect on skyrmion gyration in the free layer, the Thiele approach is considered \cite{khvalkovskiy2009vortex,everschor2012rotating}. The skyrmion is treated as a rigid profile with $\bf{X}(t)=\bf{X}[x(t), y(t)]$ representing skyrmion position. By combing the force acting on the skyrmion which is acted by the ring defect, the equation of skyrmion motion in the nanodisk can be expressed as 
\begin{equation}\label{eq:3}
\bf{G}\times\dot{\bf{X}}+\alpha \mathcal{D}\dot{\bf{X}}+\frac{\partial U}{\partial\bf{X}}+\bf{F}_{\mathrm{STT}}+\bf{F}_{\mathrm{r}}=0,
\end{equation}
where $\dot{\bf{X}}=d\bf{X}/d t$. $\bf{G}=-\bf{e}_\mathrm{z}4\pi Qt_f\mu_0M_\mathrm{s}/\gamma$ is a gyrocoupling vector, $Q=\frac{1}{4\pi}\int \mathbf{m}\cdot (\frac{\partial m}{\partial x}\times\frac{\partial m}{\partial y})$ is the skyrmion number. In our simulations, $Q=-1$, for the case that the skyrmion core magnetization orientates along -$z$-axis, and the background magnetization is along +$z$ direction. $\alpha$ is the Gilbert damping constant, and $\mathcal{D}$ is damping tensor with the tensor element $\mathcal{D}_{ij} \ (i,j=x \ \mathrm{or} \ y)$. The third term represents the force acting on skyrmion due to the nanodisk edge effect with potential $U$. $\mathbf{F}_\mathrm{STT}$ is the spin transfer force which depends on the current density flowing point-contact electrode area. $\mathbf{F}_r$ is the force from due to the interaction between ring defect and skyrmion. Due to the symmetry of the nanodisk and ring defect, the polar coordinates are considered and $d\bm{X}/dt=v_te_\theta+v_\rho e\rho$, where $v_t$ and $v_\rho$ are tangential and radial skyrmion velocity, respectively. Similarly, $\bm{F}_\mathrm{STT}$ can also be written as $\bm{F}_\mathrm{STT}=F_\mathrm{STT}^te_\theta+F_\mathrm{STT}^\rho e_\rho$, and $\bm{F}_\mathrm{r}=F_\mathrm{r}^te_\theta+F_\mathrm{r}^\rho e_\rho$. The potential $U$ is symmetric about $z$ axis, which is $U=U(\rho)$. Thus, Eq.~\ref{eq:3} can be written as
\begin{equation}
\begin{aligned}
-Gv_t+\alpha\mathcal{D}v_\rho+\frac{\partial U}{\partial \rho}-F_\mathrm{STT}^\rho-F_\mathrm{r}^\rho=&0 \\
Gv_\rho+\alpha\mathcal{D}v_t-F_\mathrm{STT}^t-F_\mathrm{r}^t=&0.
\end{aligned}
\end{equation}
When the skyrmion reaching a stable gyration in the nanodisk, the tangential velocity $v_t=\omega\rho$ and the radial velocity $v_\rho=0$, we obtain
\begin{equation}\label{eq:5}
\omega=\frac{({\partial U}/{\partial \rho}-F_\mathrm{r}^\rho)-F_\mathrm{STT}^\rho}{G\rho},
\end{equation}
or
\begin{equation}
\omega=\frac{F_\mathrm{r}^t+F_\mathrm{STT}^t}{\alpha\mathcal{D}\rho},
\end{equation}
where $\rho$ equals to the skyrmion gyration radius. Since the current density and electrode area are fixed, as well as the nanodisk size, $F_\mathrm{STT}$ and $\partial U/\partial \rho$ keep unchanged. Thus, skyrmion rotation frequency depends on $F_\mathrm{r}^\rho$, $\mathcal{D}$ and $\rho$. Consequently, introducing a ring defect modifies skyrmion rotation radius and skyrmion-defect interaction. In the following discussion, we analysis the force $F_\mathrm{r}^\rho$ which depends on the overlapping parts between defect ring and skyrmion. 
 
\begin{figure}[ht!]
	\begin{center}
		\includegraphics[width=8cm]{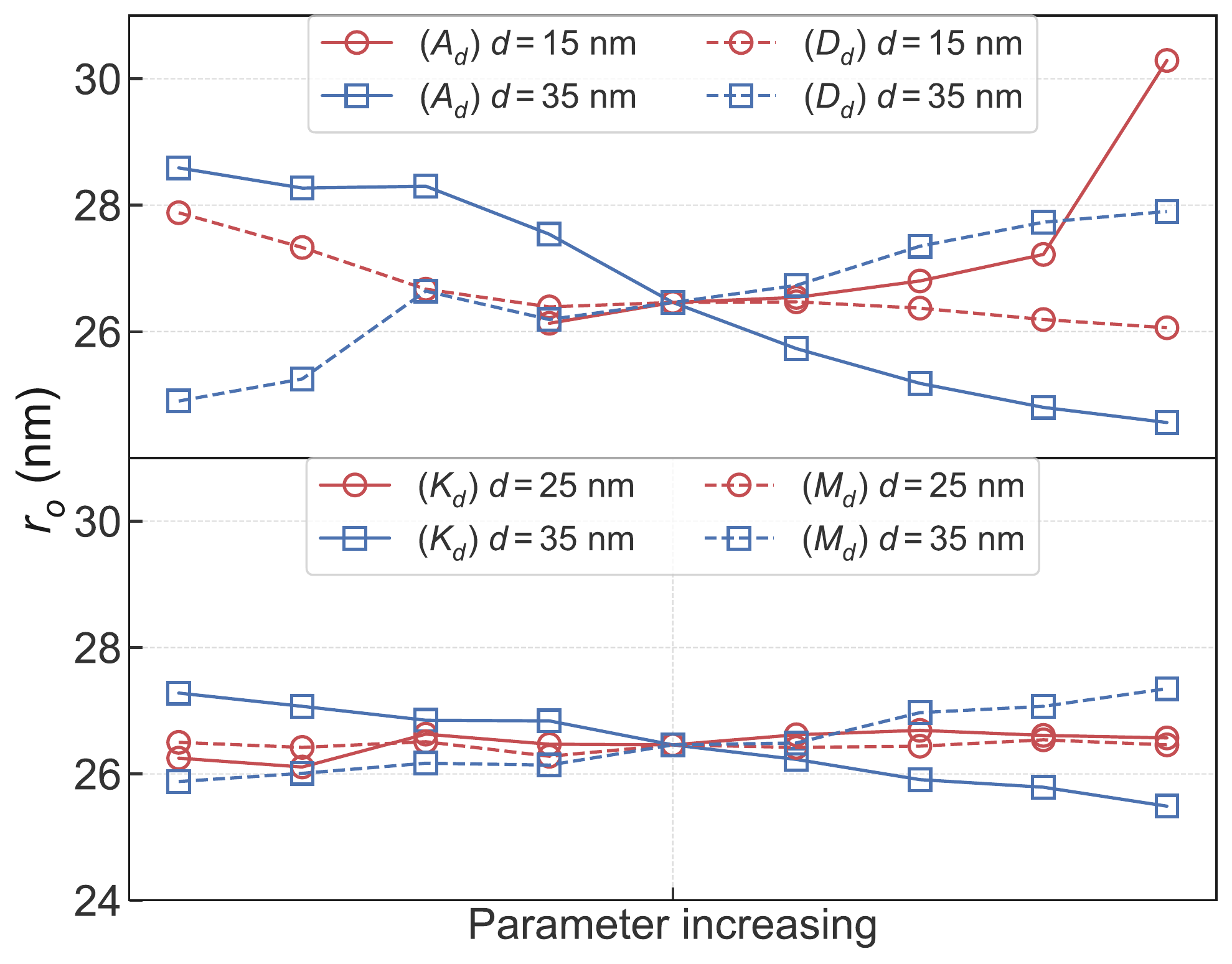} \caption{Skyrmion oscillation radius with increasing the magnetic parameters of defect ring at different $d$ at 15, 35 nm for defect with $A_d$ ($D_d$) and 25, 35 nm for defect with $K_d$ ($M_d$).}
		\label{fig9}
	\end{center}{}
\end{figure}  
In the presence of a ring defect, a potential well or barrier are formed in the nanodisk, the skyrmion moving in nanodisk will be attracted or excluded depending on the type of defect induced by the variation of magnetic parameters and $d$. Figure.~\ref{fig9} shows the skyrmion rotation radius ($r_o$) variation with increasing magnetic parameters of defect ring area at different $d$. In the presence of exchange and DMI type defect, $r_o$ decreases and increases at $d=15$ nm, respectively. As a general effect, the local increasing of $A_d$ and $D_d$ induce skyrmion size increasing (decreasing) when the skyrmion approaches attractive (repulsive) ring defect. Depending on the relative position between skyrmion and defect, at $d=15$ nm, it is repulsive for increasing $A_d$ for that the skyrmion preferred position is outside skyrmion. The corresponding potential ${\partial U}/{\partial \rho}-F_\mathrm{r}^\rho$ and oscillation frequency decreases. While the potential and frequency increase with an attractive force acted on skyrmion by increasing $D_d$, where the preferred position is at the skyrmion domain wall part. On the contrary, when $d=35$ nm, the force is repulsive with increasing $A_d$ and attractive for $D_d$, and the corresponding frequencies increases and decreases, respectively. In the case of anisotropy type defect ($K_d$), $r_o$ keeps a slight variation as increasing magnetic parameter at $d=25$ nm. The skyrmion crosses the ring defect with the outer part when rotating in the nanodisk, the overlapping part is preferred at skyrmion center. A repulsive force applied on skyrmion with increasing $K_d$, which oriented from center to edge. Nevertheless, in the case of $d=35$ nm, the skyrmion does not cross the ring defect, thus the repulsive force is oriented to center. The corresponding $r_o$ decreases with increasing $K_d$, the potential ${\partial U}/{\partial \rho}-F_\mathrm{r}^\rho$ in Eq.~\ref{eq:5} increases as well as the oscillation frequency (Fig.~\ref{fig7} (c)). For the defect with increasing magnetization ($M_d$), $r_o$ increases which is opposite to the case of $K_d$ change, and the corresponding frequency decreases as well, where a repulsive force is formed with decreasing magnetization and frequency increases. 

\section{Conclusion}
In summary, we have presented the effect of point and ring defect on the stabilization and dynamics of an isolated skyrmion in the nanodisk. We considered four different types defect with local variation of magnetic material parameters, including exchange, DMI, anisotropy and magnetization. In the presence of a point defect, the skyrmion shows a displacement due to the anisotropy attraction or repulsion. Depending on the type of parameter variation and the distance between skyrmion core and point defect, exchange and DMI defects show same influence on stabilized position of skyrmion, and a local variation of anisotropy and magnetization show same trend as well. Due to the circular symmetry of ring defect, the skyrmion shows no displacement and its size is modified, which increases or decreases depending on the type of parameter variation and inner radius of ring. We also demonstrated that the interaction of a skyrmion with ring defect leads to a large variation of rotation frequency. Our findings provide a understanding of interaction between skyrmion and defect in confined geometry, and provide a manipulation method of skyrmion by artificial engineering defect in practical applications, which is beneficial for the addressability in the stabilization of skyrmion and provide a guidance for the design of skyrmion based spin transfer nano-oscillators.   

\section*{ACKNOWLEDGMENTS}
This work is supported by National Science Fund of China (51771086). C.S. acknowledges the funding of the
China Scholarship Council.

\bibliography{reference}

\end{document}